# CTRAMER: An open-source software package for correlating interfacial charge transfer rate constants with donor/acceptor geometries in organic photovoltaic materials

*Jacob Tinnin[1,2], Huseyin Aksu[3,4], Zhengqing Tong[5], Pengzhi Zhang[1], Eitan Geva[6\*], Barry D. Dunietz[3\*], Xiang Sun[5\*], Margaret S. Cheung[1,2\*]*

[1] Department of Physics, University of Houston, 617 Science & Research Building 1, Houston, Texas 77204, USA

[2] Center for Theoretical Biological Physics, Rice University, 6500 Main St., BioScience Research Collaborative, Suite 1005G, Houston, Texas 77030-1402, USA

[3] Department of Chemistry and Biochemistry, Kent State University, 1175 Risman Drive, Kent, Ohio, 44242, USA

[4] Department of Physics, Canakkale Onsekiz Mart University, Canakkale 17100, Turkey

[5] Division of Arts and Sciences, NYU Shanghai, 1555 Century Avenue, Shanghai 200122, China; NYU-ECNU Center for Computational Chemistry at NYU Shanghai, 3663 Zhongshan Road North, Shanghai, 200062, China; Department of Chemistry, New York University, New York, New York 10003, USA

[6] Department of Chemistry, University of Michigan, 930 North University Avenue, Ann Arbor, Michigan 48109, USA

**Corresponding Authors**

*E-mail: eitan@umich.edu.

*E-mail: bdunietz@kent.edu.

*E-mail: xiang.sun@nyu.edu.

*E-mail: margaret.cheung@pnnl.gov

# Abstract

In this paper we present CTRAMER (Charge-Transfer RAtes from Molecular dynamics, Electronic structure, and Rate theory) – an open-source software package for calculating interfacial charge-transfer (CT) rate constants in organic photovoltaic (OPV) materials based on *ab-initio* calculations and molecular dynamics simulations. The software is based on identifying representative donor/acceptor geometries within interfacial structures obtained from molecular dynamics simulation of donor/acceptor blends and calculating the corresponding Fermi's golden rule CT rate constants within the framework of the linearized-semiclassical approximation. While the methods used are well-established, the integration of these state-of-the-art ideas from different disciplines to study photoinduced CT between excited states and explicit environment, in our opinion, makes this package unique and innovative. The software also provides tools for plotting other observables of interest. After outlining the features and implementation details, usage and performance of the software are demonstrated with results from an example OPV system.

# I. Introduction

The low production and environmental costs as well as improved plasticity and synthetic tunability of organic materials in comparison to their inorganic counterparts motivates the development of photovoltaic devices based on organic materials (so called organic photovoltaics [OPV]).[1-13] Within OPV devices, photoexcitation of the donor material leads to the formation of excitons. The excitons then diffuse to the donor/acceptor (D/A) interface, where charge transfer (CT), namely electron transfer from donor to acceptor, occurs.[1-3,14] This is followed by charge separation, where the electron and hole diffuse away from the interface within the donor and acceptor layers, respectively. Thus, a better understanding of the correlation between the effect of varying interfacial D/A pair geometries on CT rate constants is required to improve OPV performance.[1,6,15-18]

In this paper, we introduce a new software package, CTRAMER (CT RAtes from Molecular dynamics, Electronic structure, and Rate theory), that provides computational tools for correlating interfacial CT rates with the underlying interfacial structure. The approach, which has been benchmarked and employed by us in previous work,[1-19] combines state-of-the-art electronic structure calculations and molecular dynamics (MD) simulations to compute representative interfacial D/A geometries and the corresponding CT rate constants. The CT rate constants are calculated within the framework of Fermi's golden rule (FGR) and based on the linearized semiclassical (LSC) approximation.[19-26] Support for other levels of CT theory is planned to be added in future versions. Each of the methods used in CTRAMER have been chosen due to being well-studied and performing well in benchmarks. It is the combination of these state-of-the-art methods from different fields that makes CTRAMER unique.

It should be noted that the FGR/LSC framework currently used for calculating CT rate constants in CTRAMER is based on treating the environment of the D/A pair at the molecular level, as opposed to treatments that treat it as a polarizable-continuum or a harmonic bath.[1,21]

Such resolution is required to account for the heterogeneity of the solid state environment and the distribution of D/A geometries and CT rates it can give rise to.[1,15,20,27-31] It should also be noted that the molecular models we use are parameterized based on inputs obtained from electronic structure calculations.

Our primary goal in this paper is to introduce a general-purpose software package based on the computational framework outlined above, outline the features available in it, and demonstrate its applicability and scope by presenting results from an example OPV system where boron subphthalocyanine (SubPC) serves as the donor and fullerene ($C_{60}$) serves as the acceptor.[2,4,6,12,32-38] This framework could be broadly applied to other materials.

# II. Theory

The overall workflow for CTRAMER is outlined in Figure 1. The algorithm is initiated by the molecular coordinates of the donor and acceptor [Figure 1(a)]. The output corresponds to CT rate constants for different interfacial D/A geometries, which can be used to correlate the interfacial CT rate constants to the interfacial structure.

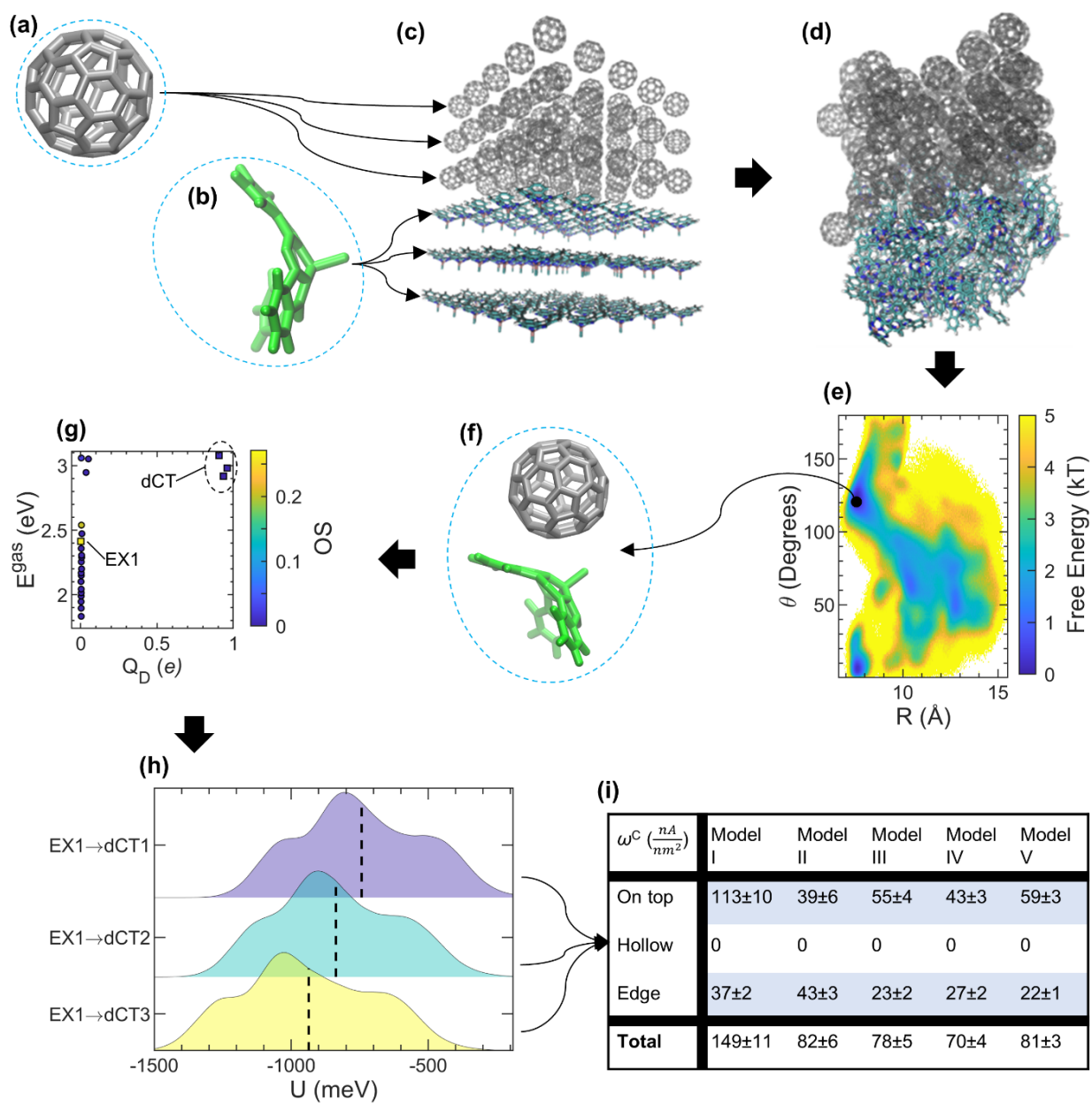

**Figure 1**. The overall workflow for CTRAMER (illustrated on the SubPC/C$_{60}$ system). **(a)/(b)** correspond to the C$_{60}$/SubPC molecules respectively, where the blue dashed enclosures represent electronic-structure calculations done for each molecule separately. **(c)** A condensed-phase system constructed using the molecular coordinates from (a) and (b) as well as the Mulliken charges obtained therein. **(d)** An equilibrated snapshot of the (c) system after equilibration. **(e)** The potential of mean force for the system in (d), where a coordinate in ($R,\theta$) space corresponds to a SubPC/C$_{60}$ pair. **(f)** A SubPC/C$_{60}$ pair selected from (e), the blue dashed enclosure representing electronic-structure calculations done on the pair. **(g)** Scatterplot of characteristics for excited states calculated in (f). **(h)** Distribution of fluctuations in the energy gap of transitions between states selected from those calculated in (f). **(i)** Charge-transfer rate-constant densities for the configuration from (f)-(h) (in this case, hollow) as well as two others. The numbers for each model vary according to the population of each configuration.

### A. Softwarization flowchart

CTRAMER consists of five modules as shown in Figure 2. The modules address the different scales needed for describing the CT process occurring at a D/A core that is affected by its molecular environment. The modular nature of CTRAMER allows the user to exclude or replace steps as needed. To initiate the calculation the individual donor and acceptor molecules [see Figures 1(a) and 1(b)] are obtained from literature or experiment. In Module 1 the atomic charges of the donor and acceptor molecules are calculated. (Alternatively, the charges can be assigned using published force-fields.) Module 2 combines multiple donor and acceptor molecules [Figure 1(c)] and equilibrates the overall system using the charges from Module 1 to a pre-assigned temperature and pressure [Figure 1(d)]. This is followed by a determination of the distribution of interfacial D/A pair geometries in a form of multiple parallel trajectories [Figure 1(e)]. Module 3 performs electronic state calculations on selected representative interfacial D/A pairs that correspond to different classes of geometries [Figure 1(f)]. From these calculations, excited donor and acceptor states are identified using a preset criteria of excited states properties (oscillator strength and CT characteristics) [Figure 1(g)]. Here the relevant transitions and the coupling between the donor and acceptor states are obtained. Module 4 uses MD simulations to calculate the fluctuations in the D/A energy gap in the condensed phase [Figure 1(h)]. Combining these fluctuations with CT characteristics from Module 3, LSC-based E-FGR CT rates [Figure 1(i)] are obtained by Module 5 at the chosen level of theory.

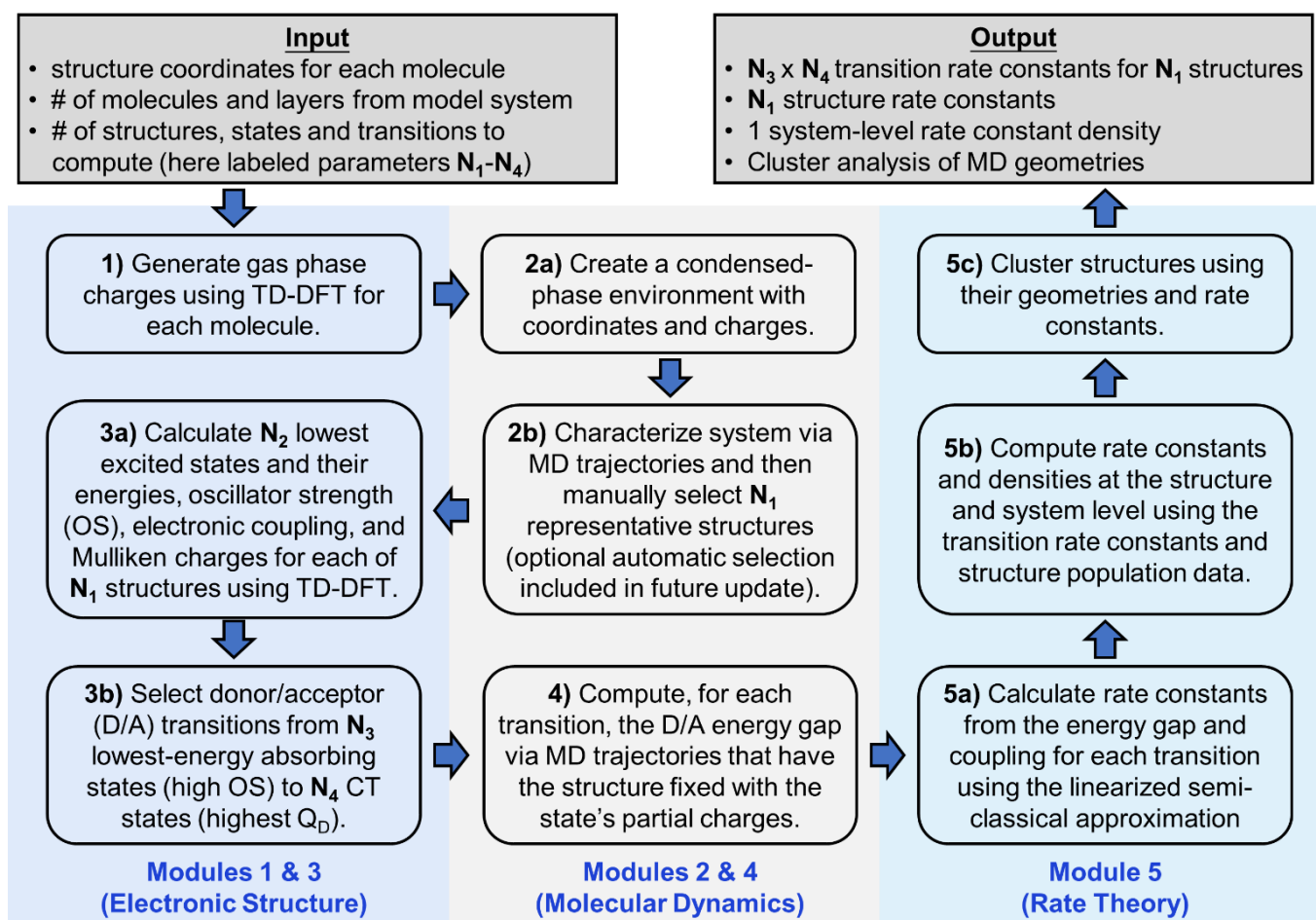

**Figure 2.** A flowchart representing information flow during each section of the software. Each box represents a process while the arrows represent data within the software. Input parameters are the number of representative structures from MD ($N_1$), excited states from Q-Chem ($N_2$), and excitonic ($N_3$) / CT ($N_4$) states for which transitions are selected.

**B. Computational Approach**

Figure 3 illustrates the actual scripts that comprise each module in CTRAMER. The electronic structure calculations and the MD simulations reported in this manuscript are based on Q-Chem 4.0[39] and AMBER12[40] respectively. All the scripts are available on Github[41] (https://github.com/ctramer/ctramer). Below, we provide additional information on the procedures for establishing the different parameters under each module.

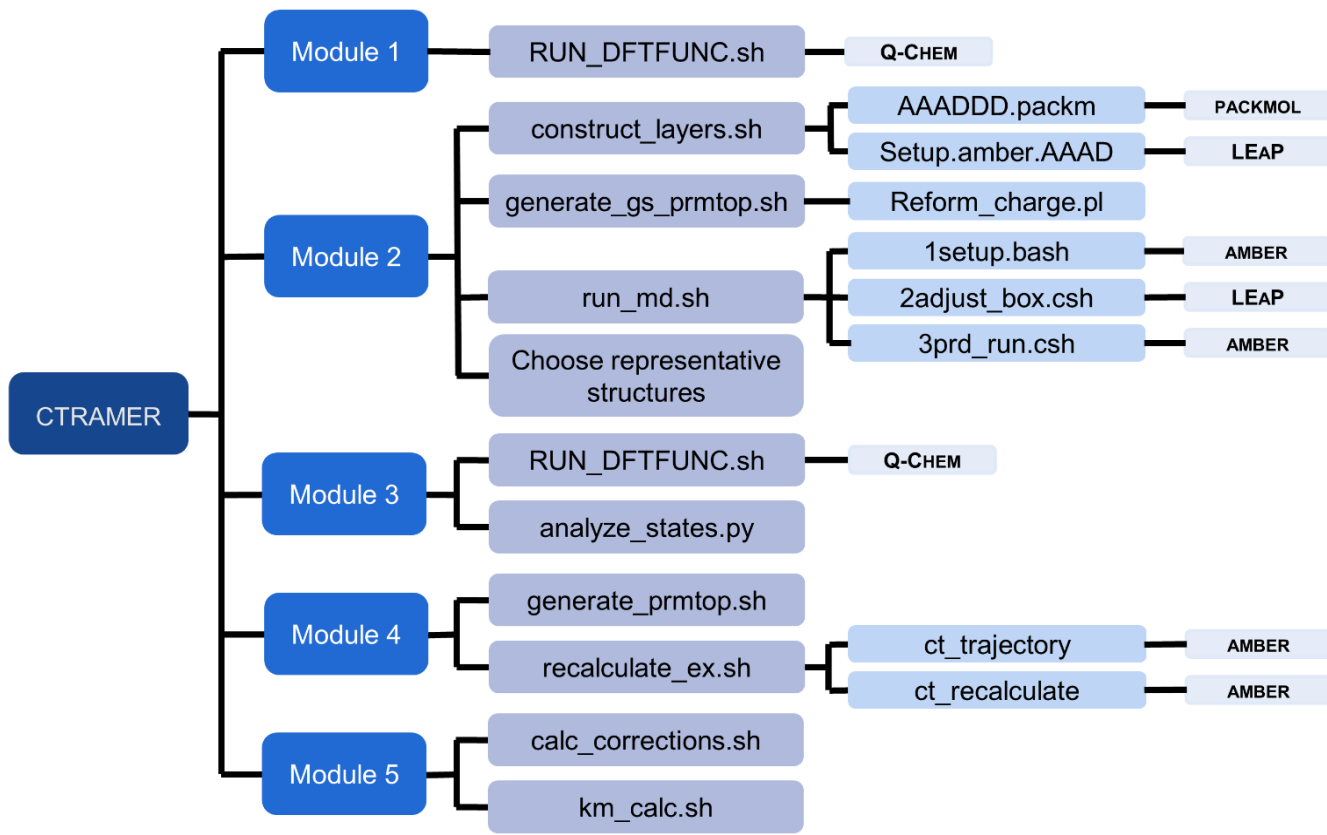

**Figure 3**. The execution of the scripts in order (available on Github: https://github.com/ctramer/ctramer).

**C. Electronic structure calculations (Modules 1 and 3)**

Time-dependent density-functional theory (TD-DFT) calculations are used in two modules of CTRAMER. First, in Module 1, TD-DFT is used on individual molecules to obtain partial charges for use in the MD simulations of Module 2. In Module 3, TD-DFT is used on selected interfacial D/A molecule pairs to obtain excited states and their partial charges, oscillator strength (OS), and relative energy.[42] These electronic structure protocols are benchmarked against experimental measured excitation energies[28,43,44] (including those of charge transfer states) as well as measured rates.[2,45] We note that, in the condensed phase, as molecules tend to neighbor several molecules, each molecule can be involved in multiple D/A pairs. We hypothesize that, on the macroscopical scale, CT follows the paths involving the D/A pairs with the fastest rate constants on the microscopical scale.

Important excited states are selected and then classified as donor and/or acceptor states. CTRAMER calculates the electronic coupling coefficients between electronic states using the fragment-charge-differences (FCD) method.[46] Choice of functional and basis set can be customized for the system under study based on literature and experimentation.

For the results presented in this paper, CTRAMER uses the 6-31G* basis set[47] and Baer-Neuhauser-Livshits (BNL) range-separated hybrid functional[48,49] for TDDFT calculations. A $\gamma$ value of 0.167 bohr$^{-1}$, tuned to an optimal value for the on-top geometry based on the $J2(\gamma)$ scheme[50], is used for all the geometries.

Calculated excited states are classified as follows. First, the charge of the donor molecule, $Q_D$, is used to classify states either as non-CT ($Q_D < 0.25\ e$) or CT ($Q_D > 0.25\ e$). Second, states with a significant OS are referred to as light-absorbing states, or *bright*.[51] CT states with negligible OS are referred to as *dark* (dCT) and those with a significant OS are addressed as bright CT (bCT). Non-CT states with significant OS are referred to as excitonic (EX). As CTRAMER studies direct CT from photoexcited (bright) states to CT states, states that are both dark and non-CT are not addressed. However, extensions to consider additional processes can be developed in future versions. The states are then named as EXn, bCTn, or dCTn. The index *n* refers to the rank of a state's energy, from smallest to largest, within EX, bCT, and dCT states with the same geometry. Classification thresholds can be customized within CTRAMER.

Last, using these classified states, transitions from a bright state to a CT state are selected. The user selects the maximum number of donor and acceptor states between which transition rates are calculated. The computational cost is linear with the number of donor states but is largely unaffected by the number of acceptor states (due to the method of calculating rates detailed in the following subsection). As part of these electronic structure calculations, electronic coupling coefficients are computed for each transition.

**D. Molecular dynamics simulations (Modules 2 and 4)**

Module 2 focuses on creating a condensed-phase system from the coordinates and charges established in Module 1, equilibrating it to a given temperature and pressure, and then analyzing the distribution of structures. From this distribution, representative structures are selected for further analysis along with their surrounding molecules. In simulations during Module 4, an interfacial molecular pair is kept fixed for use in calculating energy gaps between excited states. These energy gaps are used in Module 5 to calculate rate constants.

CTRAMER uses Packmol[52] to construct condensed-phase OPV systems. Packmol places a user-defined number of molecules into a region of space, the size of which is also set by the user. CTRAMER by default uses 6 square layers into each of which it places 25 molecules. Each layer consists of only one type of molecule. Space is placed between separate types of layers to simulate the fabrication procedure. Too much or too little space between molecules can cause the system to not equilibrate correctly.

The MD simulations of CTRAMER utilize the generalized AMBER force field (GAFF),[40,53,54] augmented by other force fields (FFs) as needed. For example, interactions involving the boron atom that is found at the center of SubPC are not given in GAFF and are here taken from Refs. [55,56] The MD simulations are performed by the AMBER12 program SANDER.[40] CTRAMER fits the simulation box to the constructed system and then applies periodic boundary conditions. Both equilibration and production MD simulations have a time step of 2.0 fs. The SHAKE algorithm[57] is used to constrain bonds involving hydrogen. Neighbor list updates, real-space Coulomb interactions, and van der Waals interactions utilize a 10.0 Å threshold. CTRAMER calculates electrostatic interactions with the particle-mesh Ewald method.[58,59]

By default, a hybrid algorithm is used to perform energy minimization. The default settings for this algorithm, used for the results in this paper, are listed here. The conjugated-gradient method is used for 500 timesteps, followed by 4500 timesteps of the steepest-descent method. The minimized system is then heated to 298.15K using the canonical ensemble (NVT) gradually over a period of 10 ps. CTRAMER then fixes the system at this temperature for 1.0 ns and equilibrates at a constant pressure of 1.0 bar using a NPT ensemble with 1.0 ps as the pressure relaxation time. The simulation box is then refit to the equilibrated system with periodic boundary conditions kept. A planned future update is for CTRAMER to simulate automatically until equilibration is reached.

Production runs in Module 2 use the equilibrated system with the adjusted box size in a NVT ensemble. As with all other parameters, the number and length of production runs can be specified by the user. For the results shown in this paper, one run of 20 ns was used. In our tests on CTRAMER, multiple production runs can aid in sampling multiple local minima.

Productions run in Module 4 use the entire system in the timestep in Module 2 from which a representative geometry was chosen. The atoms of the representative pair are fixed in place using a harmonic potential. By default, CTRAMER uses a force constant of 50 kcal/mol-$\text{Å}^2$. However, as the energies calculated by CTRAMER do not include the restraint potential, the size of the force constant will not affect the results.[40,60] The goal of fixing the atoms is to keep the molecules in the representative geometry. These fixed atoms are assigned with Mulliken partial charges, from Module 3, of a donor state for that representative pair geometry. The rest of the molecules in the system have ground state charges from Module 1 and are unrestrained. After equilibration, production runs are performed using the same parameters as Module 2. The effect on uncertainty in rate constants in Module 5 by changing the number and length of production runs in Module 4 is discussed in the following section.

**E. Rate evaluations (Module 5)**

Module 5 evaluates CT rates at different levels. Transition rate constants are calculated following a Marcus-level linearized semi-classical (LSC) [1, 19, 21, 22] approximation to Fermi's Golden Rule, where the donor-to-acceptor transition rate constant ($k^M$) is given by:

$$k^M = \frac{|\Gamma_{DA}|^2}{\hbar}\sqrt{\frac{2\pi}{\sigma_U^2}}\exp\left[-\frac{\langle U\rangle^2}{2\sigma_U^2}\right]. \quad (1)$$

Here, $U(\mathbf{R}) = V_D(\mathbf{R}) - V_A(\mathbf{R})$ is the D/A energy gap as a function of the nuclear coordinates, $\mathbf{R}$, where $V_D$ and $V_A$ are the potential-energy surfaces of the donor and acceptor states, respectively. $\langle U\rangle$ and $\sigma_U$ are the first and second moments, respectively, of $U(\mathbf{R})$. This rate constant is derived from Fermi's Golden Rule by the LSC approximation, as detailed in refs. [19,21,23,24], but is referred to as "Marcus-level" as the reorganization energy, $E_r$, the reaction free energy, $\Delta E$, and the activation energy, $E_a$, can be calculated from $\langle U\rangle$ and $\sigma_U$: $E_r = \sigma_U^2/(2k_BT)$, $\Delta E = -E_r - \langle U\rangle$, and $E_a = k_BT\langle U\rangle^2/(2\sigma_U^2)$. These parameters allow for the analysis of rates calculated using Marcus Theory.[61-64] The electronic excitation energies are from the Module 3 electronic structure calculations.

$U(\mathbf{R})$ is determined for each transition using the classically sampled trajectories from Module 4 as detailed in the MD methods section. First, the potential energy surface from MD at a given timestep, $V_\alpha^M$, is calculated for each state, with α denoting donor or acceptor, by recalculating the energy of the entire system. To avoid double counting of potential energy by the electronic structure and MD methods, $V_\alpha^M$ is corrected to $V_\alpha = V_\alpha^M + W_\alpha$, where $W_\alpha$ is the difference in single-point energy of each state between calculations by the electronic structure and MD methods. $U$ is then determined using the difference of $V_D$ and $V_A$. Finally, $\langle U\rangle$ and $\sigma_U$ are obtained using the moments of $N_R$ MD trajectories of length $L_R$. We here use $N_R$=40 runs and $L_R$=40 ns, which can be adjusted for the desired accuracy. The uncertainty for rate constants decreases approximately as $\frac{1}{\sqrt{N_R L_R}}$.

As shown in Figure 1(h) and Figure S1, the distribution of $U(\mathbf{R})$ can be significantly non-Gaussian for many transitions. We attribute this to CTRAMER accounting for how the condensed phase's heterogeneity can lead to multiple local energy minima. As the LSC approximation uses a Gaussian distribution to model the probability density at $U(\mathbf{R}) = 0$, CTRAMER resolves these energy minima by best fitting the probability density function of $U(\mathbf{R})$ to a sum of Gaussian distributions chosen by least-squares regression. CTRAMER increases the number of distributions, beginning at one, until the 95% confidence interval for $\sigma_U$ includes non-positive numbers or a maximum of three is reached. These settings can be customized within CTRAMER.

While k$^M$ measures the transition rate between states, we multiply it by the amount of the corresponding charge transferred, ($\Delta Q_D$), to measure the rate of CT (k$^c$) for a transition:

$$k^c = \Delta Q_D \, k^M. \tag{2}$$

Both k$^c$ and k$^M$ are obtained in the context of a single transition for one structure. For aid in comparison between structures, these CT rates are summed over all the identified transitions, $t$, for a given representative structure $i$, to give a structure-level CT rate constant ($K_i^C$):

$$K_i^C = \sum_t k_t^C. \tag{3}$$

A system-level CT density ($\omega^C$) can then established by averaging K$^c$ over the area of the D/A interface in the simulation:

$$\omega^C = \sum_i (K_i^C \, n_i) \times \frac{1}{A}, \tag{4}$$

where $n_i$ is the number of pairs represented by the structure $i$ from Equation 3 and A is the approximate area of the interface (which is by default calculated by CTRAMER but can replaced).

# III. Results

CTRAMER is used to analyze the CT rate in the interface of SubPC/$C_{60}$ pair of donor acceptor organic material used in model OPV studies. CTRAMER can be used for many materials but these results are presented here as an example use of the software. Data is available at https://github.com/ctramer/ctramer.[41]

### A. Sample preparation (Module 1)

The coordinates of the optimized SubPC and $C_{60}$ molecules are provided in Github[41] (https://github.com/ctramer/ctramer), while the references used to determine the atomic charges are listed in Table S1.

### B. Condensed-phase structures (Module 2)

The multilayered OPV system is represented using 6 alternating layers of 25 SubPC or $C_{60}$ molecules each. A large ensemble of interfacial D/A pairs is then obtained, where a interfacial D/A pair is defined as 1 SubPC molecule and 1 $C_{60}$ molecule where the minimum distance between any atom from separate molecules is less than 5 Å.

Analysis of the different interfacial pairs is aided by two order parameters [see Figure 4(b)]: first, the distance ($R$) between the SubPC boron atom and the $C_{60}$ center of mass, and second, the angle ($\theta$) between the vector from the $C_{60}$ center of mass to the SubPC boron atom and the vector from the SubPC boron atom to the SubPC chlorine atom.

Inspection of the potential of mean force (PMF) using a ($R,\theta$) coordinate system [shown in Figure 4(a)] reveals two pronounced regions of low energy centered at approximately (7.5 Å, 0º) and (7.5 Å, 120º), which correspond to D/A pair geometries identified as *on top* and *hollow* in previous gas-phase and mean-field studies[1,2,33,50,65] [examples of which are shown in Figure 4(a)]. The approximate percentage of sampled D/A pairs corresponding to these geometries [shown in Figure 4(a)] are calculated using D/A pairs where $R < 8.5$ Å and $\theta < 38°$ for *on top* and D/A pairs where $R < 9$ Å and $95° < \theta < 160°$ for *hollow*.

However, the majority of the D/A pairs in the PMF does not correspond to either *on top* or *hollow*. Instead, these pairs correspond to a geometry, noted as *edge*, that was recently identified using a condensed-phase analysis.[1] While most of the *edge* pairs fall within the range of 10 Å $< R <$ 14 Å and $40° < \theta < 100°$, *they* have a large variance in $R$ and $\theta$ [one example structure is shown in Figure 4(a)]. The *edge* geometry consists of pairs where only the *edge* of a SubPC arm is close to the $C_{60}$ molecule. Most of the *edge* pairs fall within the following range on the PMF: 10 Å $< R <$ 14 Å and $40° < \theta < 100°$

Among the ensemble structures from each basin (e.g. *on-top*, *hollow*, or *edge* ensembles), the most probable structure is selected as the representative D/A pair geometry for analysis in Module 2.

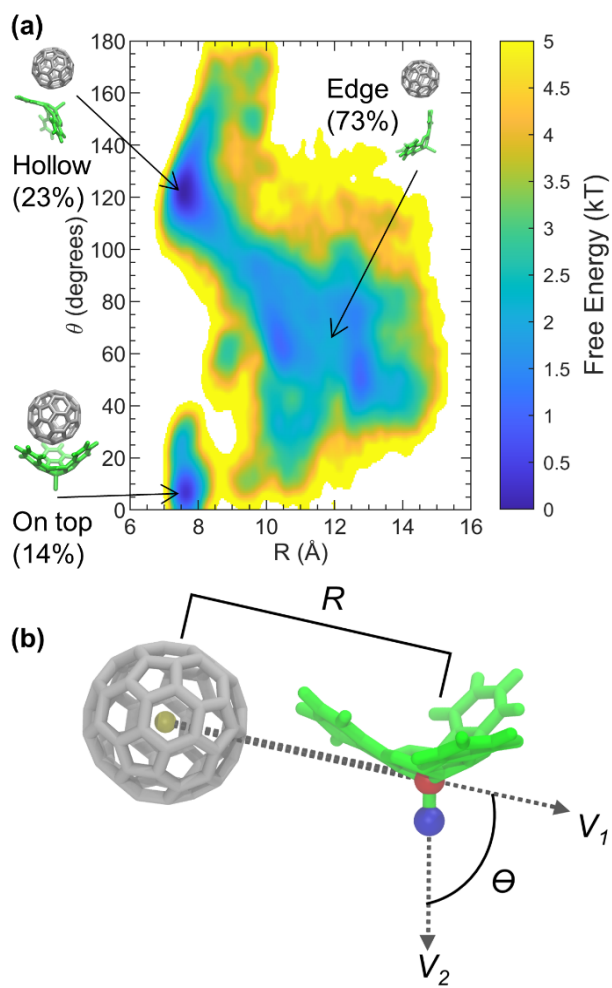

**Figure 4.** The potential of mean force (a) for the SubPC/C$_{60}$ pair on a $R$-$\theta$ coordinate system. The color is scaled by k$_B$T. Also shown are representative structures (*on top*, *hollow*, and *edge*) of SubPC/C$_{60}$ pair at each of the three major geometries and the percentage of interfacial pairs in that geometry. $R$ and $\theta$ are defined in (b), where the yellow bead corresponds to the C$_{60}$ center of mass and the red and blue beads correspond to the SubPC boron and SubPC chlorine atoms, respectively. $R$ is the distance between the SubPC boron atom and the C$_{60}$ center of mass. $\theta$ is the angle between vectors $V_1$ and $V_2$.

## C. Electronic structure results (Module 3)

Electronic structure calculations are performed on the interfacial D/A geometries selected from Module 2. The important parameters are (1) the charge of the D/A pair's donor molecule ($Q_D$), (2) the energy for the D/A pair while isolated in the gas-phase ($E^{gas}$), and (3) the pair's oscillator strength (OS).[51] These three parameters are shown by the x axis, y axis, and color bar, respectively, in Figure 5.

Transitions studied here are from the lowest-energy bright EX state to each of the dark CT states, with one exception: in the on-top geometry, a bright CT state is considered as a donor state in addition to the lowest-energy EX state and also as an acceptor state. The OS of the bright states for the *on-top* geometry is about half that of the EX1 state for the *hollow* and *edge* geometries, as shown in Figure 5 and listed in Table S2. *Hollow* and *edge* each also have another EX state that is similar in both OS and $E^{gas}$ but is not used for these results. However, the dCT states for *on top* range from approximately equal to slightly lower in energy than its bright states while all the dCT states for *hollow* and *edge* are much higher in energy than the bright states. The exact values shown in Figure 5 are recorded in Table S2. EX and CT states as labeled in Figure 5 are used for analysis in Module 4.

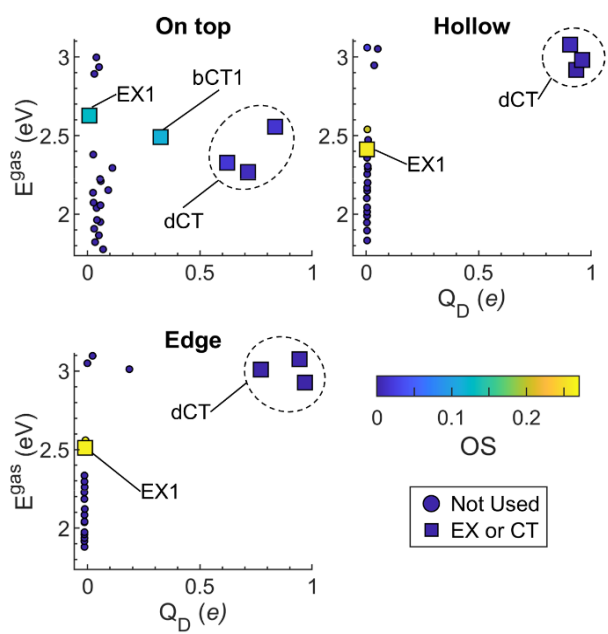

**Figure 5.** Scatter plot of the excitation energy in the gas phase ($E^{gas}$) versus the charge of the donor molecule ($Q_D$) for the 25 lowest excited states of each representative geometry. Dots are colored according to their oscillator strength (OS). States used for calculations in this paper are shown by large squares and labelled with their names while other states are denoted by smaller circles.

**D. Charge-transfer rate constants (Modules 4 and 5)**

Next, we calculate the electronic-population-transfer rate constant, $k^M$, and CT rate constant, $k^C$, for transitions from a donor state (EX1 or bCT1) to an acceptor state (bCT1, dCT1, dCT2, or dCT3). The required inputs to calculate $k^M$, as shown in Eq 1, are the D/A energy-gap first and second moments, the coupling coefficient, $\Gamma_{DA}$, and the excitation-energy correction, $W_\alpha$, (see Tables 1 and S3). The relative energies between states are changed in MD simulations due to the molecular environment. Clearly, the CT states are expected to be stabilized by the condensed-phase polarizable environment more than the localized excitations.[27] The rate constant, $k^C$, (see Table 1) is the product of $k^M$ and $\Delta Q_D$, the difference in charge of the donor molecule between the donor and acceptor states.

In the *on-top* geometry, the $\Gamma_{DA}$ for the bCT1 → dCT1 and bCT1 → dCT2 transitions are at least double the size of any other $\Gamma_{DA}$ considered. On the other hand, as transitions involving bCT1 use a CT state as donor state, $\Delta Q_D$ for these transitions is significantly less than that of transitions with an EX donor state. As a result, the values of $k^C$ for transitions with either EX1 or bCT1 as the donor state in the *on-top* geometry are within an order of magnitude of each other. Additionally, the values of $k^M$ are comparable to those from a non-condensed-phase analysis based on optimized geometries.[2]

For both the *hollow* and *edge* geometries, smaller $\Gamma_{DA}$ values for transitions from EX1 to dCT states than those in the *on-top* geometry lead to smaller rate constants. However, the $k^C$ values for the *edge* geometry are larger overall than those of the *hollow* geometry. This difference between *edge* and *hollow* can be traced back to the fact that the transitions in the edge geometry correspond to a much smaller $\langle U \rangle$ than those of the hollow geometry (see Table 1). The $k^M$ values for transitions in the *hollow* geometry are observed to be significantly smaller than those previously reported in a non-condensed-phase analysis based on optimized geometries (see Table S4).[2] Marcus theory parameters are available for all transitions in the Supplemental Information (Table S5).

**Table 1. Interfacial Charge-Transfer Rates for Representative Geometries**

| Geometry | Transition | $\Gamma_{DA}$ (meV) | $\Delta Q_D$ ($e$) | $\langle U \rangle$ (meV) | $k^C$ (nA) |
|---|---|---|---|---|---|
| *On-top* | EX1→dCT1 | 4.03 | 0.706 | 217±2 | 180±10 |
| | EX1→dCT2 | 24.46 | 0.616 | 148±11 | 520±70 |
| | EX1→dCT3 | 5.47 | 0.825 | -168±1 | 220±10 |
| | EX1→bCT1 | 25.82 | 0.314 | 114±5 | 510±40 |
| | bCT1→dCT1 | 74.16 | 0.391 | 123±2 | 24±2 |
| | bCT1→dCT2 | 72.27 | 0.302 | 61.6±0.1 | 1,010±20 |
| | bCT1→dCT3 | 21.13 | 0.511 | -240±12 | 0.014±0.008 |
| *Hollow* | EX1→dCT1 | 1.85 | 0.943 | -743±14 | 0.006±0.002 |
| | EX1→dCT2 | 20.21 | 0.959 | -837±14 | 0.04±0.02 |
| | EX1→dCT3 | 15.53 | 0.905 | -936±30 | 0.006±0.004 |
| *Edge* | EX1→dCT1 | 10.30 | 0.977 | -354±23 | 65±5 |
| | EX1→dCT2 | 14.02 | 0.781 | -432±15 | 0.32±0.07 |
| | EX1→dCT3 | 17.22 | 0.953 | -481±20 | 3.1±0.6 |

# IV. Conclusions

We have described the software package CTRAMER for the analysis of CT rates based on electronic structure calculations, MD simulations, and rate theory. CTRAMER is a unique combination of well-established methods from different disciplines that allows for a more precise study of photoinduced CT between excited states and explicit environment. The customizable features, software architecture, and guidelines for usage were discussed. Additionally, the scientific justification behind the different approaches as well as example results were described.

CTRAMER will continue to be actively developed and supported, as it will remain a long-term focus of the authors. Additions planned for the immediate future are automated tuning of the parameters for electronic structure calculations, enabling different level of theory, and machine-learning clustering methods for selecting representative structures. Other future goals include various extensions to improve the accessibility and computational efficiency of the software.

# Supplementary Material

See supplementary material for additional information regarding the results: parameters used for MD not included in GAFF, a table of excited state properties, energy correction terms for each state, information on non-Gaussian energy gap distributions, and Marcus theory parameters.

# Acknowledgements

E.G., B.D.D., and M.S.C. acknowledge support from the Department of Energy (DOE), Basic Energy Sciences through the Chemical Sciences, Geosciences and Biosciences Division (No. DE-SC0016501). X.S. acknowledges support from the National Natural Science Foundation of China (No. 21903054), and the Program for Eastern Young Scholar at Shanghai Institutions of Higher Learning. Computational resources are provided by the National Energy Research Scientific Computing Center (NERSC), a U.S. Department of Energy Office of Science User Facility operated under Contract No. DE-AC02-05CH11231, as well as the uHPC cluster managed by the University of Houston and acquired through NSF Award Number OAC 1531814.

# Data Availability

The data that support the findings of this study are openly available in Github, reference number [41].

# References

1. J. Tinnin, S. Bhandari, P. Z. Zhang, H. Aksu, B. Maiti, E. Geva, B. D. Dunietz, X. Sun and M. S. Cheung, Phys. Rev. Appl. **13** (5), 11 (2020).
2. M. H. Lee, E. Geva and B. D. Dunietz, J. Phys. Chem. C **118** (18), 9780 (2014).
3. Y. Zhao and W. Z. Liang, Chem. Soc. Rev. **41** (3), 1075 (2012).
4. J. Kim and S. Yim, Appl. Phys. Lett. **99** (19), 3 (2011).
5. H. N. Tian, Z. Yu, A. Hagfeldt, L. Kloo and L. Sun, J. Am. Chem. Soc. **133** (24), 9413 (2011).
6. X. R. Tong, B. E. Lassiter and S. R. Forrest, Org. Electron. **11** (4), 705 (2010).
7. A. Hagfeldt, G. Boschloo, L. C. Sun, L. Kloo and H. Pettersson, Chemical Reviews **110** (11), 6595 (2010).
8. T. H. Wang and J. L. Bredas, Matter **2** (1), 119 (2020).
9. J. Kurpiers, T. Ferron, S. Roland, M. Jakoby, T. Thiede, F. Jaiser, S. Albrecht, S. Janietz, B. A. Collins, I. A. Howard and D. Neher, Nature Communications **9** (1), 2038 (2018).
10. S. M. Falke, C. A. Rozzi, D. Brida, M. Maiuri, M. Amato, E. Sommer, A. De Sio, A. Rubio, G. Cerullo, E. Molinari and C. Lienau, Science **344** (6187), 1001 (2014).
11. L. Benatto, C. F. N. Marchiori, T. Talka, M. Aramini, N. A. D. Yamamoto, S. Huotari, L. S. Roman and M. Koehler, Thin Solid Films **697**, 10 (2020).
12. N. C. Giebink, G. P. Wiederrecht, M. R. Wasielewski and S. R. Forrest, Physical Review B **82** (15), 155305 (2010).
13. N. C. Giebink, G. P. Wiederrecht, M. R. Wasielewski and S. R. Forrest, Physical Review B **83** (19), 195326 (2011).
14. M. H. Lee, B. D. Dunietz and E. Geva, J. Phys. Chem. Lett. **5** (21), 3810 (2014).
15. D. Balamurugan, A. J. A. Aquino, F. de Dios, L. Flores, H. Lischka and M. S. Cheung, J. Phys. Chem. B **117** (40), 12065 (2013).
16. J.-L. Brédas, D. Beljonne, V. Coropceanu and J. Cornil, Chemical Reviews **104** (11), 4971 (2004).
17. V. D'Innocenzo, G. Grancini, M. J. P. Alcocer, A. R. S. Kandada, S. D. Stranks, M. M. Lee, G. Lanzani, H. J. Snaith and A. Petrozza, Nature Communications **5** (1), 3586 (2014).
18. C. W. Tang and S. A. VanSlyke, Appl. Phys. Lett. (51), 913 (1987).
19. X. Sun and E. Geva, J. Chem. Theory Comput. **12** (6), 2926 (2016).
20. X. Sun, P. Z. Zhang, Y. F. Lai, K. L. Williams, M. S. Cheung, B. D. Dunietz and E. Geva, J. Phys. Chem. C **122** (21), 11288 (2018).
21. X. Sun and E. Geva, J. Phys. Chem. A **120** (19), 2976 (2016).
22. Z. Hu, Z. Tong, M. S. Cheung, B. D. Dunietz, E. Geva and X. Sun, J. Phys. Chem. B **124** (43), 9579 (2020).
23. X. Sun and E. Geva, J. Chem. Phys. **144** (24), 12 (2016).
24. X. Sun and E. Geva, J. Chem. Phys. **145** (6), 19 (2016).
25. Z. Tong, X. Gao, M. S. Cheung, B. D. Dunietz, E. Geva and X. Sun, J. Chem. Phys. **153** (4), 9 (2020).
26. X. Sun, P. Zhang, Y. Lai, K. L. Williams, M. S. Cheung, B. D. Dunietz and E. Geva, The Journal of Physical Chemistry C **122** (21), 11288 (2018).
27. A. K. Manna, D. Balamurugan, M. S. Cheung and B. D. Dunietz, J. Phys. Chem. Lett. **6** (7), 1231 (2015).

28. H. Aksu, A. Schubert, E. Geva and B. D. Dunietz, J. Phys. Chem. B **123** (42), 8970 (2019).
29. M. A. Steffen, K. Q. Lao and S. G. Boxer, Science **264** (5160), 810 (1994).
30. A. Niedringhaus, V. R. Policht, R. Sechrist, A. Konar, P. D. Laible, D. F. Bocian, D. Holten, C. Kirmaier and J. P. Ogilvie, Proc. Natl. Acad. Sci. U. S. A. **115** (14), 3563 (2018).
31. J. Koepke, E. M. Krammer, A. R. Klingen, P. Sebban, G. M. Ullmann and G. Fritzsch, J. Mol. Biol. **371** (2), 396 (2007).
32. K. L. Mutolo, E. I. Mayo, B. P. Rand, S. R. Forrest and M. E. Thompson, J. Am. Chem. Soc. **128** (25), 8108 (2006).
33. N. Ilyas, S. S. Harivyasi, P. Zahl, R. Cortes, O. T. Hofmann, P. Sutter, E. Zojer and O. L. A. Monti, J. Phys. Chem. C **120** (13), 7113 (2016).
34. G. D'Avino, L. Muccioli, F. Castet, C. Poelking, D. Andrienko, Z. G. Soos, J. Cornil and D. Beljonne, Journal of Physics: Condensed Matter **28** (43), 433002 (2016).
35. D. S. Josey, S. R. Nyikos, R. K. Garner, A. Dovijarski, J. S. Castrucci, J. M. Wang, G. J. Evans and T. P. Bender, ACS Energy Letters **2** (3), 726 (2017).
36. R. Pandey, A. A. Gunawan, K. A. Mkhoyan and R. J. Holmes, Advanced Functional Materials **22** (3), 617 (2012).
37. D. E. Wilcox, M. Lee, M. E. Sykes, A. Niedringhaus, E. Geva, B. Dunietz, M. Shtein and J. Ogilvie, The journal of physical chemistry letters **6 3**, 569 (2015).
38. M. H. Lee, E. Geva and B. D. Dunietz, J. Phys. Chem. A **120** (19), 2970 (2016).
39. Y. Shao, L. F. Molnar, Y. Jung, J. Kussmann, C. Ochsenfeld, S. T. Brown, A. T. B. Gilbert, L. V. Slipchenko, S. V. Levchenko, D. P. O'Neill, R. A. DiStasio, R. C. Lochan, T. Wang, G. J. O. Beran, N. A. Besley, J. M. Herbert, C. Y. Lin, T. Van Voorhis, S. H. Chien, A. Sodt, R. P. Steele, V. A. Rassolov, P. E. Maslen, P. P. Korambath, R. D. Adamson, B. Austin, J. Baker, E. F. C. Byrd, H. Dachsel, R. J. Doerksen, A. Dreuw, B. D. Dunietz, A. D. Dutoi, T. R. Furlani, S. R. Gwaltney, A. Heyden, S. Hirata, C. P. Hsu, G. Kedziora, R. Z. Khalliulin, P. Klunzinger, A. M. Lee, M. S. Lee, W. Liang, I. Lotan, N. Nair, B. Peters, E. I. Proynov, P. A. Pieniazek, Y. M. Rhee, J. Ritchie, E. Rosta, C. D. Sherrill, A. C. Simmonett, J. E. Subotnik, H. L. Woodcock, W. Zhang, A. T. Bell, A. K. Chakraborty, D. M. Chipman, F. J. Keil, A. Warshel, W. J. Hehre, H. F. Schaefer, J. Kong, A. I. Krylov, P. M. W. Gill and M. Head-Gordon, Physical Chemistry Chemical Physics **8** (27), 3172 (2006).
40. D. A. Case, T. A. Darden, I. T.E. Cheatham, C. L. Simmerling, J. Wang, R. E. Duke, R. Luo, R. C. Walker, W. Zhang, K. M. Merz, B. Roberts, S. Hayik, A. Roitberg, G. Seabra, J. Swails, A. W. Götz, I. Kolossváry, K. F. Wong, F. Paesani, J. Vanicek, R. M. Wolf, J. Liu, X. Wu, S. R. Brozell, T. Steinbrecher, H. Gohlke, Q. Cai, X. Ye, J. Wang, M.-J. Hsieh, G. Cui, D. R. Roe, D. H. Mathews, M. G. Seetin, R. Salomon-Ferrer, C. Sagui, V. Babin, T. Luchko, S. Gusarov, A. Kovalenko and P. A. Kollman, (University of California, San Francisco, 2012).
41. J. Tinnin, H. Aksu, Z. Tong, P. Z. Zhang, E. Geva, B. D. Dunietz, X. Sun and M. S. Cheung, 2021, CTRAMER SubPC/C60 Data, Github, https://github.com/ctramer/ctramer
42. E. Runge and E. K. U. Gross, Phys. Rev. Lett. **52** (12), 997 (1984).
43. K. Begam, S. Bhandari, B. Maiti and B. D. Dunietz, J. Chem. Theory Comput. **16** (5), 3287 (2020).
44. S. Bhandari and B. D. Dunietz, J. Chem. Theory Comput. **15** (8), 4305 (2019).
45. M. H. Lee, B. D. Dunietz and E. Geva, J. Phys. Chem. C **117** (44), 23391 (2013).
46. A. A. Voityuk and N. Rosch, J. Chem. Phys. **117** (12), 5607 (2002).
47. V. A. Rassolov, J. A. Pople, M. A. Ratner and T. L. Windus, J. Chem. Phys. **109** (4), 1223 (1998).
48. R. Baer and D. Neuhauser, Phys. Rev. Lett. **94** (4), 4 (2005).
49. E. Livshits and R. Baer, Physical Chemistry Chemical Physics **9** (23), 2932 (2007).

50. T. Stein, L. Kronik and R. Baer, J. Chem. Phys. **131** (24), 5 (2009).
51. J. W. Robinson, *Atomic Spectroscopy*, 2nd ed. (Marcel Dekker, Inc., New York, 1996).
52. L. Martinez, R. Andrade, E. G. Birgin and J. M. Martinez, J. Comput. Chem. **30** (13), 2157 (2009).
53. J. M. Wang, W. Wang, P. A. Kollman and D. A. Case, J. Mol. Graph. **25** (2), 247 (2006).
54. J. M. Wang, R. M. Wolf, J. W. Caldwell, P. A. Kollman and D. A. Case, J. Comput. Chem. **25** (9), 1157 (2004).
55. C. G. Claessens, D. González-Rodríguez and T. Torres, Chemical Reviews **102** (3), 835 (2002).
56. X. Wu, Z. Liu, S. Huang and W. Wang, Physical Chemistry Chemical Physics **7** (14), 2771 (2005).
57. G. Ciccotti and J. P. Ryckaert, Computer Physics Reports **4** (6), 345 (1986).
58. T. Darden, D. York and L. Pedersen, J. Chem. Phys. **98** (12), 10089 (1993).
59. D. M. York, T. A. Darden and L. G. Pedersen, The Journal of Chemical Physics **99** (10), 8345 (1993).
60. J. Han, P. Z. Zhang, H. Aksu, B. Maiti, X. Sun, E. Geva, B. D. Dunietz and M. S. Cheung, J. Chem. Theory Comput. **16** (10), 6481 (2020).
61. P. F. Barbara, T. J. Meyer and M. A. Ratner, J. Phys. Chem. **100** (31), 13148 (1996).
62. R. A. Marcus, J. Chem. Phys. **24** (5), 966 (1956).
63. R. A. Marcus, J. Chem. Phys. **24** (5), 979 (1956).
64. R. A. Marcus, Rev. Mod. Phys. **65** (3), 599 (1993).
65. C. Xu, G. Li and S. Zheng, Journal of Photochemistry and Photobiology A: Chemistry **403**, 112852 (2020).